\documentclass[twocolumn,english,twocolumn]{revtex4}
\usepackage[T1]{fontenc}
\usepackage[latin9]{inputenc}
\usepackage{amstext}
\usepackage{amssymb}
\usepackage{graphicx}

\makeatletter

\newcommand{\noun}[1]{\textsc{#1}}

\@ifundefined{textcolor}{}
{%
 \definecolor{BLACK}{gray}{0}
 \definecolor{WHITE}{gray}{1}
 \definecolor{RED}{rgb}{1,0,0}
 \definecolor{GREEN}{rgb}{0,1,0}
 \definecolor{BLUE}{rgb}{0,0,1}
 \definecolor{CYAN}{cmyk}{1,0,0,0}
 \definecolor{MAGENTA}{cmyk}{0,1,0,0}
 \definecolor{YELLOW}{cmyk}{0,0,1,0}
 }

\makeatother

\usepackage{babel}
\begin{document}

\title{Accumulation of beneficial mutations in one dimension}

\author{Jakub Otwinowski}

\email{jotwinowski@physics.emory.edu}

\affiliation{Physics Department, Emory University, Atlanta, Georgia 30322}

\author{Stefan Boettcher}

\email{sboettc@emory.edu}

\affiliation{Physics Department, Emory University, Atlanta, Georgia 30322}
\begin{abstract}
When beneficial mutations are relatively common, competition between
multiple unfixed mutations can reduce the rate of fixation in well-mixed
asexual populations. We introduce a one-dimensional model with a steady
accumulation of beneficial mutations. We find a transition between
periodic selection and multiple-mutation regimes. In the multiple-mutation
regime, the increase of fitness along the lattice bears a striking
similarity to surface growth phenomena, with power law growth and
saturation of the interface width. We also find significant differences
compared to the well-mixed model. In our lattice model, the transition
between regimes happens at a much lower mutation rate due to slower
fixation times in one dimension. Also the rate of fixation is reduced
with increasing mutation rate due to the more intense competition,
and it saturates with large population size. 
\end{abstract}
\maketitle

\section{Introduction}

In population genetics, the study of the rate at which mutations are
generated and incorporated into populations has been largely a theoretical
activity due to the difficulty of measuring changes in organisms over
many generations. The simplest case is called \noun{periodic selection},
when beneficial mutations are rare such that each mutation has ample
time to spread to the whole population before the next mutation arrives.
We assume conditions such that harmful mutations die out quickly and
survive at a negligible rate, so we only consider beneficial mutations.
In this regime the rate of fixation is limited by the rate at which
mutations appear in the population. 

\begin{figure}
\begin{centering}
\includegraphics{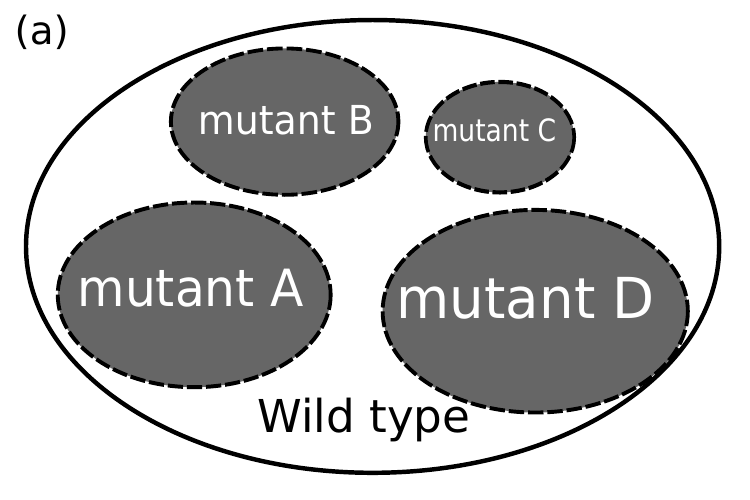}
\par\end{centering}

\begin{centering}
\includegraphics{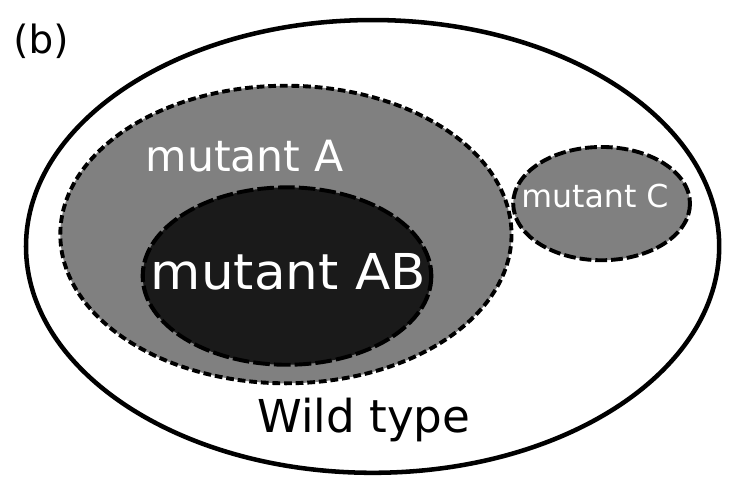}
\par\end{centering}

\caption{\label{fig:Clonal-Interference}All mutations shown are assumed to
be beneficial compared to the wild-type. (a) Clonal interference:
mutation A has to compete with mutations B, C, and D, reducing its
chances of fixation compared to the case when there is only one mutation.
If mutation A fixates, mutations B, C, and D are {}``wasted'', slowing
down the rate of fixation . (b) Multiple-mutation effect: Mutation
AB arises in a population with Mutation A, increasing mutation A's
chances of fixation.}
\end{figure}

However, recent experiments in microbes suggest that beneficial mutations
may be more common than previously thought \citep{Sniegowski2010}.
The competition between multiple unfixed beneficial mutations is termed
\emph{clonal interference }\citep{Gerrish1998} (Fig \ref{fig:Clonal-Interference}a).
In this picture some good mutations must be wasted because only one
of them ultimately fixates, which reduces the average probability
of fixation, and reduces the rate of fixation. Larger mutations are
more likely to survive competition, eliminating mutations of weak
effect, and biasing the distribution of fixated mutational effects.
Sexual organisms may alleviate this problem by recombining mutations,
known as the \noun{Fisher-Muller hypothesis} \citep{Muller1932,Fisher1930},
An analogue of clonal interference when mutations are linked is known
as the \noun{Hill-Robertson effect} \citep{Hill1966,Barton1995}.
Since our goal is to focus on the effects of spatial structure, we
avoid complications associated with sexual reproduction and study
only asexual organisms.

The original clonal interference model neglected the possibility that
an individual may acquire multiple beneficial mutations. Assuming
mutations are additive, an additional mutation improves the chances
of fixation of the first mutation instead of suppressing it, and the
simultaneous fixation of multiple mutations becomes possible (Fig
\ref{fig:Clonal-Interference}b) \citep{Kim2005,Barrett2006,Desai2007,Desai2007a,Fogle2008}.
Current research concerns a complete description that takes into account
both aspects \citep{Hallatschek2010a} (for a review see \citep{Sniegowski2010,Park2010}).
Generally, research in clonal interference and multiple mutations
find the rate of adaptation, or speed of evolution, to be slower than
periodic selection. Clonal interference analysis finds fixation to
occur in isolated instances, as in the periodic selection regime.
In contrast, multiple mutation analysis finds fixation occurs in clusters,
and because of the stacking of mutations, no mutation ever dominates
the entire population. Simulation studies have found that the relative
importance of either effect depends on the distribution of beneficial
mutations effects \citep{Fogle2008,Park2007}. If the distribution
has a heavy tail, and large effects are relatively common, then clonal
interference can dominate. If the distribution has a sharp cutoff,
and large effects are relatively rare, then effects from multiple
mutations can be more important. Unfortunately, we do not know what
distributions are found in nature, and it may depend on the level
of adaptation to the environment \citep{Orr2010,Martin2008}.

In this paper we will investigate the accumulation of beneficial mutations
of a population with spatial structure. While the fixation probability
of a mutant on a spatially structured population is usually the same
as in a well-mixed population \citep{Maruyama1970,Lieberman2005},
the time scales can be much slower \citep{ChavesFilho2010,Whitlock2003,Habets2007}.
In a well-mixed population every individual competes with each other,
but with spatial structure the spread of a mutant is restricted by
space. Nothing happens inside a domain where the fitnesses are all
the same. The only changes are at the boundaries which are defined
by fitness differences. Gordo and Campos \citep{Gordo2006} studied
the speed of evolution on a 2D lattice. They found the speed to be
slower, and the time to fixation to be longer, than in a well-mixed
population, and their results were supported by experiments with bacteria
\citep{Perfeito2008}. Others have studied the loss of genetic variation
in 1D stepping stone models to describe the boundary of an expanding
bacterial colony \citep{Hallatschek2007,Hallatschek2008,Hallatschek2010,Korolev2010}.
Starting with multiple alleles, they found that over time the population
segregates into domains of single alleles. The effects of drift and
selection change significantly, since they act only on the domain
boundaries. The timescale of segregation was found to be slower than
in the well-mixed case (algebraic instead of exponential). 

We chose to study a one-dimensional spatial structure because it is
the simplest structure where we would expect the most deviation from
well-mixed models. While Hallatschek and Nelson also studied the accumulation
of beneficial mutations in an expanding frontier in the non-interacting
regime \citep{Hallatschek2010}, we will study a model where mutations
are common enough to interfere with each other. In section \ref{sec:model},
we will introduce a Wright-Fisher model on a 1D lattice and study
the dependence on the rate of beneficial mutations, and the size of
the population. Our model has three timescales that are not well separated:
selection, mutation and stochasticity (drift). Such three-timescale
models are difficult to analyze analytically and we must resort to
simulations for most of our insights. In section \ref{sec:surfacegrowth}
we discuss the similarity to surface growth phenomena, and in section
\ref{sec:Discussion} we discuss how our results remain qualitatively
the same under more general conditions.

\section{\label{sec:model}1-D model with mutations and selection}

Our model consists of a 1-D lattice with $N$ sites and periodic boundary
conditions. Unlike stepping stone models with sub-populations or demes,
there is only one asexual haploid individual at every site. Time is
discrete and represents each generation (parallel updates), and the
total population $N$ stays constant. Each generation dies and is
replaced by its offspring which inherit the fitness of their single
parent. The major change from well-mixed models is that we specify
a spatial neighborhood which limits where parents may have their children.
For simplicity we chose the smallest possible neighborhood of size
2. An organism at site $i$ and time $t$ may have children at time
$t+1$ at sites $i$ and $i+1$ when $t$ is odd and sites $i$ and
$i-1$ when $t$ is even. The new generation is chosen so the number
of children of each parent is proportional to its fitness relative
to its neighbors' fitness. In simulation this amounts to each child
{}``choosing'' its parent weighted by their fitnesses. For each
child at site $i$, the fitness is copied from parent site $i$ with
probability $f_{i}/(f_{i}+f_{i\pm1})$ or from site $i\pm1$ with
probability $ $$f_{i\pm1}/(f_{i}+f_{i\pm1})$. 

Every generation, the number of mutations is determined by a Poisson
random number with mean $UN$. The effect of a mutation is to increase
the fitness, $f_{i}$, multiplicatively as $f'_{i}=f_{i}(1+s)$, where
$s$ is a small constant %
\footnote{In general, $s$ may be drawn from a distribution, and this may have
a significant effect on the dynamics as discussed above. Since we
do not know the true distribution, we have some freedom in choosing
one. Desai et al argued that $s$ should have a characteristic value
because clonal interference eliminates smaller $s$, and large $s$
are unlikely to begin with \citep{Desai2007,Desai2007a}.%
}. In the multiplicative fitness model the fitness of a new mutation
relative to its neighbors remains the same since common factors in
the fitness drop out. The balance of mutation and selection leads
to a steady state in the variance of the log fitnesses. The fitness
averaged over the population, $\bar{f}(t)$ will increase exponentially
as a function of time, and the \emph{speed of evolution} is the rate
constant defined as: 
\begin{equation}
v=\lim_{t\rightarrow\infty}\frac{\langle\ln\bar{f}(t)\rangle}{t},
\end{equation}
where brackets indicate the \emph{ensemble} average. Since $s$ is
constant, $v$ is related to the rate of fixation $R$, simply as
$R=v/s$. 

\begin{figure}
\begin{centering}
\includegraphics[width=1\columnwidth]{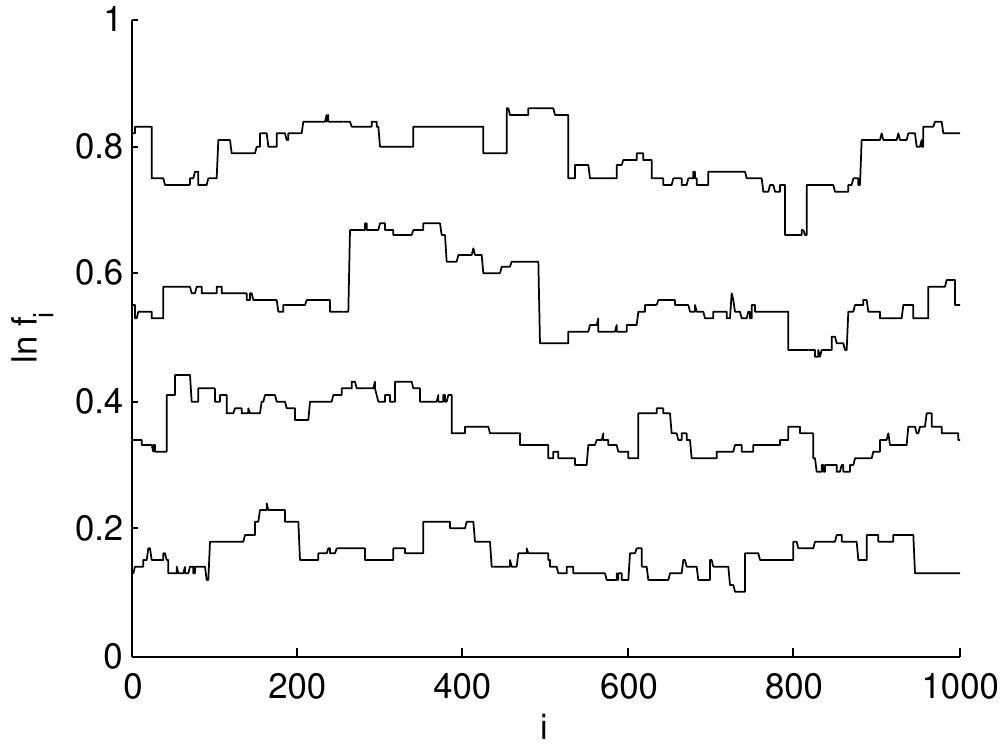}
\par\end{centering}

\caption{\label{fig:front}The logarithm of the fitness varies in spatial position.
This interface  moves with velocity $v$ and has a standard deviation
$\sigma$. $N=1000$, $U=10^{-3}$, $s=0.01$. Shown are snapshots
separated by 10000 generations}
\end{figure}

In the competing mutation regime, after an initial transient phase
a progressing interface forms in log-fitness space reminiscent of
surface growth phenomena (figure \ref{fig:front}).
 The interface of the logarithm of the fitness has a distribution
that fits extremely well to a traveling Gaussian wave with a speed
$v$ and a standard deviation $\sigma$. In general, these quantities
depend on $U$ and $N$. The additive increases in log-fitness and
spreading of mutations parallel the addition of particles and smoothing
of the interface in surface growth. From the simulation it was determined
that the system reaches a steady state after $\sim100N$ generations.

As in the original multiple-mutations model, the population has a
moving distribution of fitnesses with a steady width. This is in contrast
to the original clonal interference model where the distribution of
fitnesses varied and dropped to zero when fixation occurred. 

\begin{figure}
\begin{centering}
\includegraphics[width=1\columnwidth]{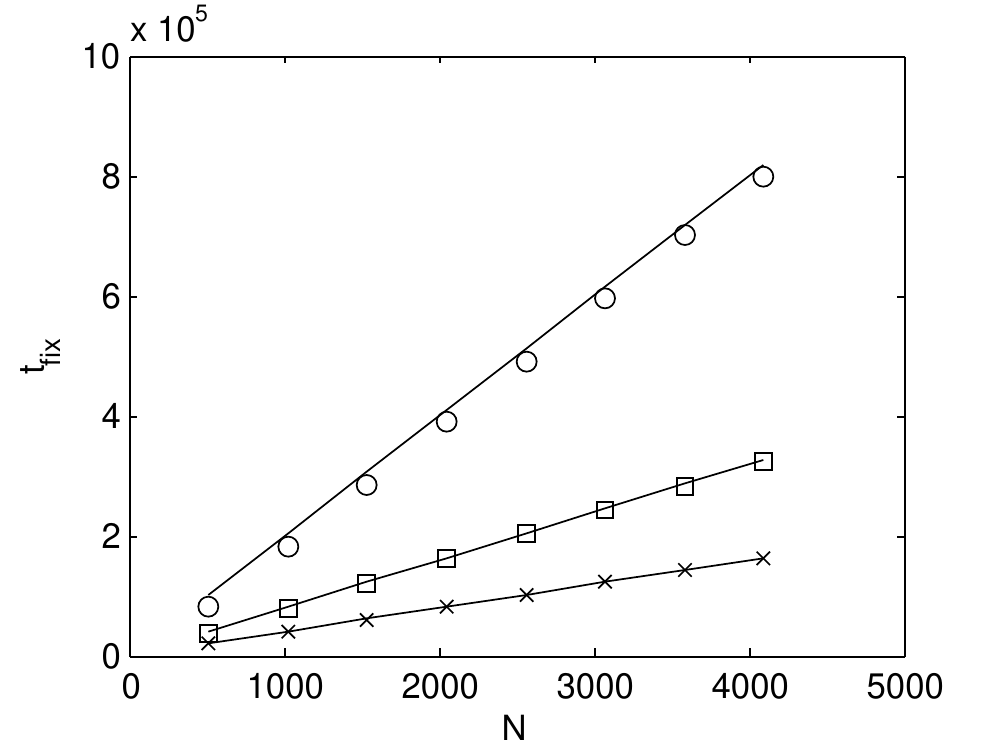}
\par\end{centering}

\caption{\label{fig:tfix}Fixation times were determined by planting single
mutations with $s=0.01$ (circles), $s=0.025$ (squares), and $s=0.05$
(x's), averaged over 100 fixations. Lines indicate $t_{fix}=2N/s$}
\end{figure}

The multiple-mutation regime occurs when the fixation time is approximately
equal to or greater than the time for mutations to appear and establish
themselves: 
\begin{equation}
t_{\textrm{fix}}\gtrsim t_{\textrm{mut}}
\end{equation}
The fixation probability, $\pi$, for a single mutation happens to
be the same as in the well-mixed model \citep{Lieberman2005}, $\pi=2s$
for large $Ns$ and small $s$ \citep{KIMURA1962}. The average time
between fixating mutations is 
\begin{equation}
t_{\textrm{mut}}\approx\frac{1}{2sUN}.\label{eq:tmut}
\end{equation}
In the periodic selection regime, each mutation has time to spread
to the whole population before the next mutation arrives, or $t_{\textrm{fix}}\ll t_{\textrm{mut}}$,
and the rate of fixation is mutation limited: 
\begin{equation}
R_{s}=\frac{1}{t_{\textrm{mut}}}=2sUN.\label{eq:v1}
\end{equation}
The transition to the multiple-mutation regime occurs when $t_{\textrm{mut}}\sim t_{\textrm{fix}}$.
The fixation time for a single mutant can be formulated as a first
passage problem. The fixation time is the mean time for a stochastic
particle (representing the size of the mutant domain) to first reach
position $N$ without ever reaching position $0$. The largest contributing
term to the fixation time (when $N$ is not too small) is simply the
size of the system divided by the drift velocity, or $ $ 
\begin{equation}
t_{\textrm{fix}}=\frac{2N}{s}.\label{eq:tfix}
\end{equation}
 This is confirmed with simulations in figure \ref{fig:tfix}, although
some deviation is present from terms of order $s^{-2}$.

\begin{figure}
\begin{centering}
\includegraphics[width=1\columnwidth]{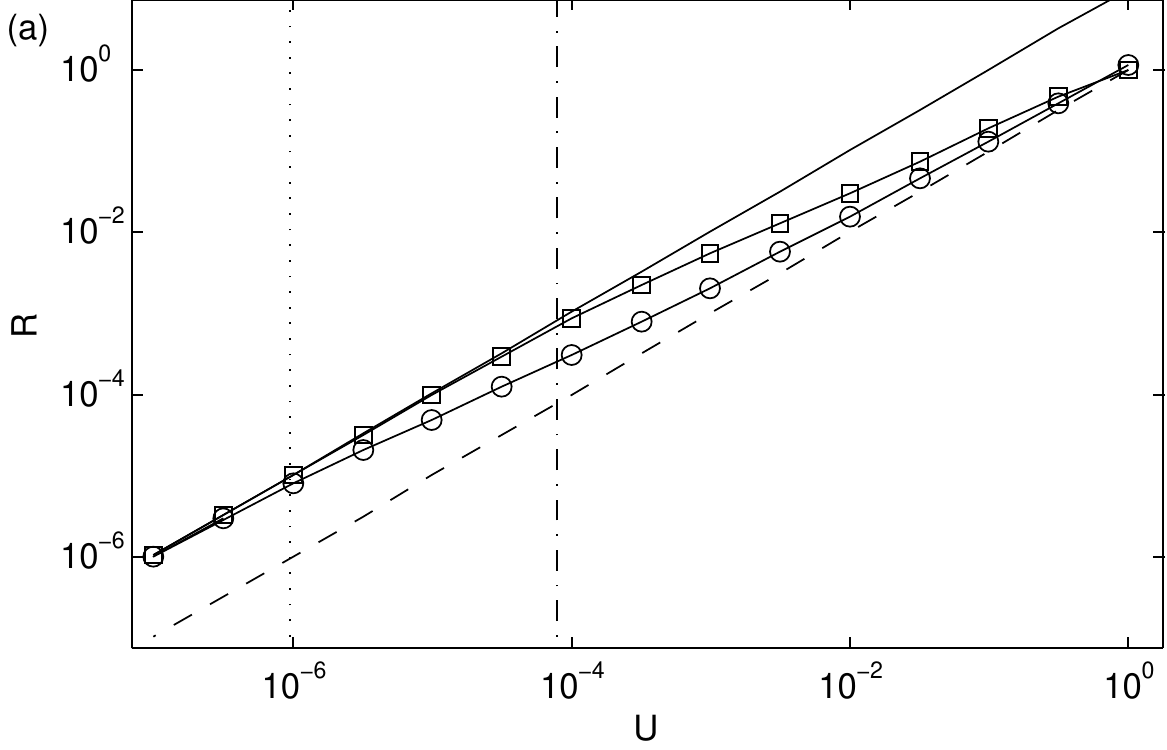}
\par\end{centering}

\begin{centering}
\includegraphics[width=1\columnwidth]{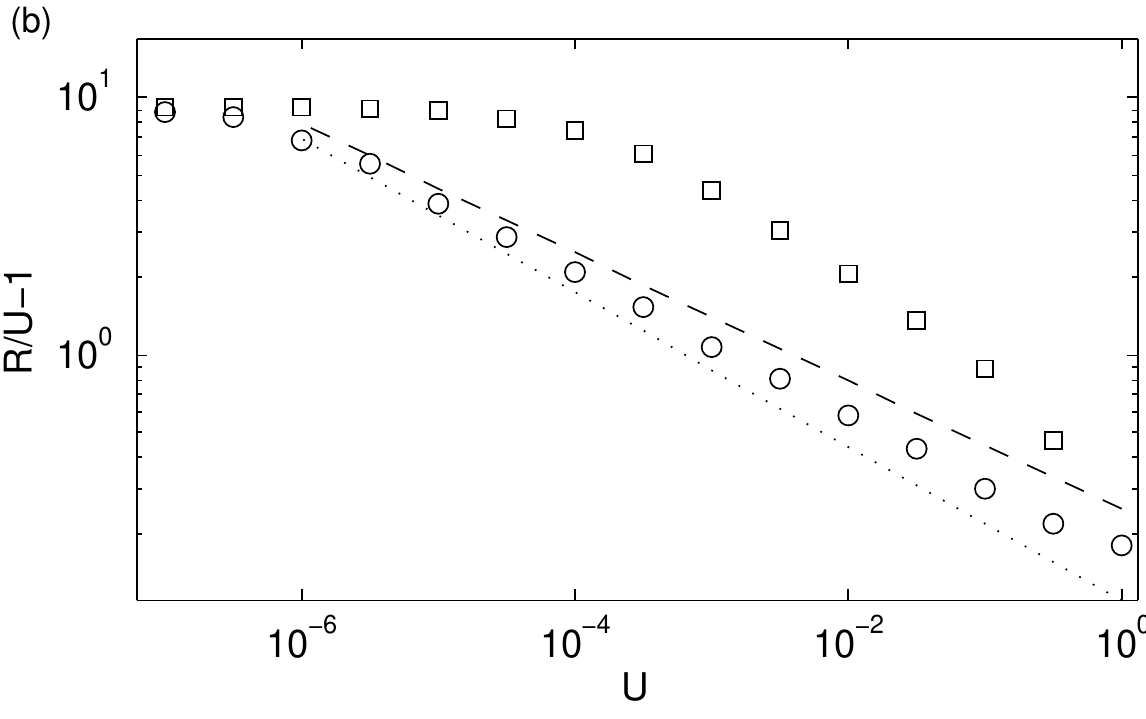}
\par\end{centering}

\caption{\label{fig:v}(a) Fixation rate $R$ versus mutation rate $U$ with
$N=512$ and $s=0.01$ averaged over $10^{3}$ fixations for the 1D
model (circles) and well-mixed model (squares). The dotted line indicates
the transition between the single-fixation regime and the multiple-mutation
regime for the spatial model, and the dot-dashed line indicates the
transition for the well-mixed wright-fisher model. The solid line
is the non-interfering fixation rate (\ref{eq:v1}), and the dashed
line is the neutral fixation rate, $R_{n}=U$. (b) The difference
between the 1D model (circles) and the well-mixed model (squares)
is clearer with $R/U-1$, which approaches zero as $R$ approaches
the neutral fixation rate $R_{n}=U$. As guides, the dashed and dotted
lines indicate power law relations with exponents -0.25, and -0.3
respectively.}
\end{figure}

The transition between the regimes is obtained by equating (\ref{eq:tmut})
and (\ref{eq:tfix}) which results in: $U_{\textrm{tr}}\sim1/(4N^{2})$.
Figure \ref{fig:v} shows that the fixation rate follows (\ref{eq:v1})
in the periodic selection regime As $U$ approaches 1, every site
receives a mutation at every generation on average, and no mutation
has a relative fitness advantage. Therefore the rate of fixation approaches
the neutral fixation rate, $R=U$ (figure \ref{fig:v}a, dashed line).
Between the two extremes we observe a transition.

 We also simulate the standard well-mixed Wright-Fisher model with
constant $s$ according to \citep{Park2010}. Since the fixation time
is $t_{\textrm{fix}}=2\ln(N)/s$ , the transition happens at $U\sim1/(4N\ln N)$.
Figure \ref{fig:v}a shows that the transition happens at much higher
$U$ in the well-mixed model, and the gap separating the two transitions
grows with system size. The difference is more clearly seen by dividing
$R$ with $U$ , shown in figure \ref{fig:v}b. The fixation rate
approaches the neutral fixation rate as a power law $ $$\sim U^{\gamma}$,
with $\gamma=0.3-0.4$ 

\begin{figure}
\begin{centering}
\includegraphics[width=1\columnwidth]{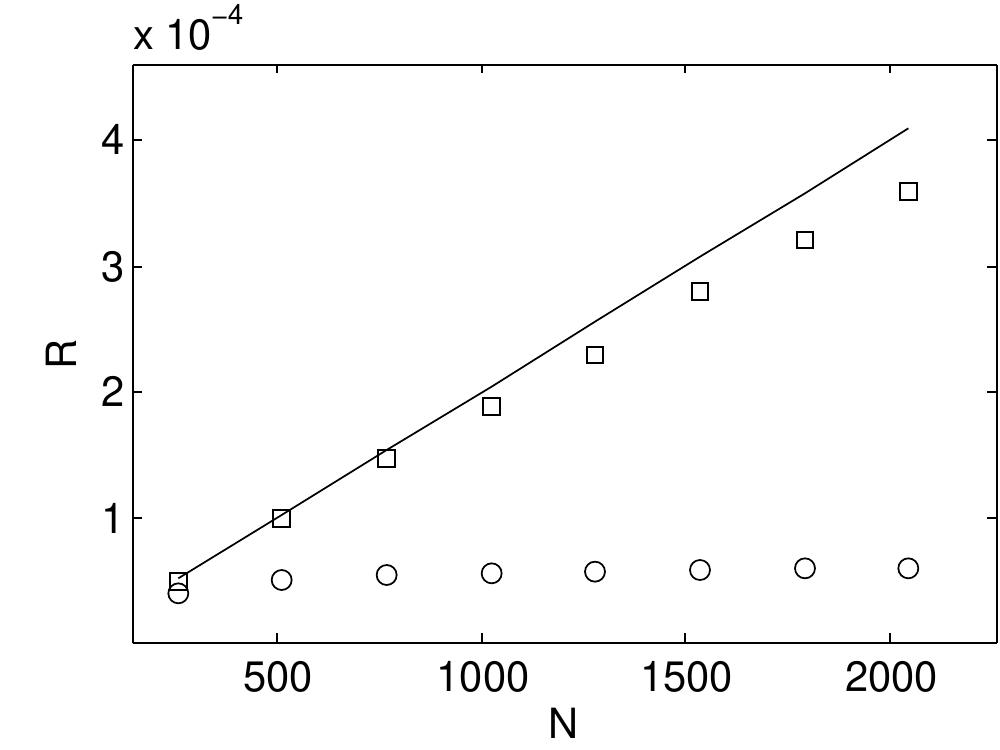}
\par\end{centering}

\caption{\label{fig:R_N}Fixation rate, $R$, versus system size for the 1D
model (circles) and the well-mixed model (squares) with $U=10^{-5}$.
$v$ quickly saturates in 1D but diverges for the well-mixed model.
Solid line is the non-interfering fixation rate (\ref{eq:v1}).}
\end{figure}

The fixation rate is always higher in the well-mixed model than in
one dimension. The difference in $R$ between the models may seem
small, however it is a logarithmic plot, and more importantly the
difference depends on the system size $N$. Figure \ref{fig:R_N}
shows that $R$ becomes independent of $N$ at large $N$, while the
well-mixed model has a sub-linear dependence on $N$. The saturation
of $R$ with system size is not intuitive. Mutations must either fixate
or be lost, so one may write down a conservation relation: 
\begin{equation}
UN=R+D,
\end{equation}
where $D$ is the loss rate. In the case where $R$ does not depend
on $N$ ($U\gg1/(4N^{2})$), the loss rate must increase with system
size to balance the increase in mutations. $D$ must take the form:
\begin{equation}
D\approx UN\left(1-\frac{A}{N}\right),
\end{equation}
 where $A$ is some function of $s$ and $U$. Solving for $A$ yields:
\begin{equation}
A=R/U,
\end{equation}
which is shown in figure \ref{fig:v}b, valid only in the multiple-mutation
regime $U\gg1/(4N^{2})$. Presumably, $A$ is a power law of $U$,
and an unknown function of $s$. 

\section{\label{sec:surfacegrowth}Similarity to Surface Growth}

It is illustrative to exploit the similarity between our model and
surface growth. In surface growth phenomena, the width or standard
deviation of the interface grows initially in time as $\sigma\sim t^{\beta}$,
where $\beta$ is the growth exponent, then reaches a steady state
when the correlation length reaches the size of the system \citep{Barabasi1995,Family1985}.
In the steady state $\sigma\sim N^{\alpha}$ where $\alpha$ is the
saturation exponent. We found the width or standard deviation of the
distribution of fitnesses also follows power-laws with critical exponents
as a function of $N$ and $U$.

\begin{figure}
\begin{centering}
\includegraphics[width=1\columnwidth]{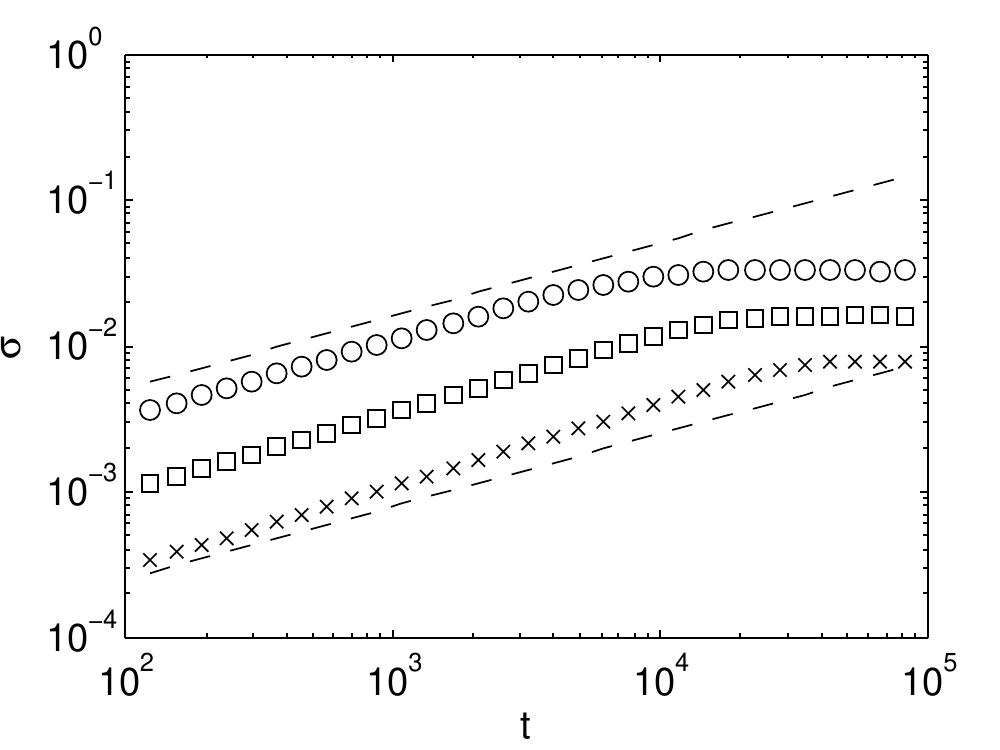}
\par\end{centering}

\caption{\label{fig:transients}Time evolution of the standard deviation of
the fitnesses $\sigma$ with $N=512$, $s=0.01$ and $U=10^{-3}$
(circles), $U=10^{-4}$ (squares), and $U=10^{-5}$ (crosses). The
dashed lines are $\sim t^{1/2}$ The width approaches a stationary
value as a power law with time $\sigma\sim t^{\beta}$, where $\beta$
is the growth exponent. $\beta$ is approximately $1/2$ for high
$U$, and increases slightly as $U$ gets smaller.}
\end{figure}

By averaging over many simulations we found the transient values of
$\sigma$ in figure \ref{fig:transients}. The width increased as
$\sigma\sim t^{\beta}$, although $\beta$ was slightly dependent
on $U$. At high $U$, $\beta$ was close to one half. 

\begin{figure}
\begin{centering}
\includegraphics[width=1\columnwidth]{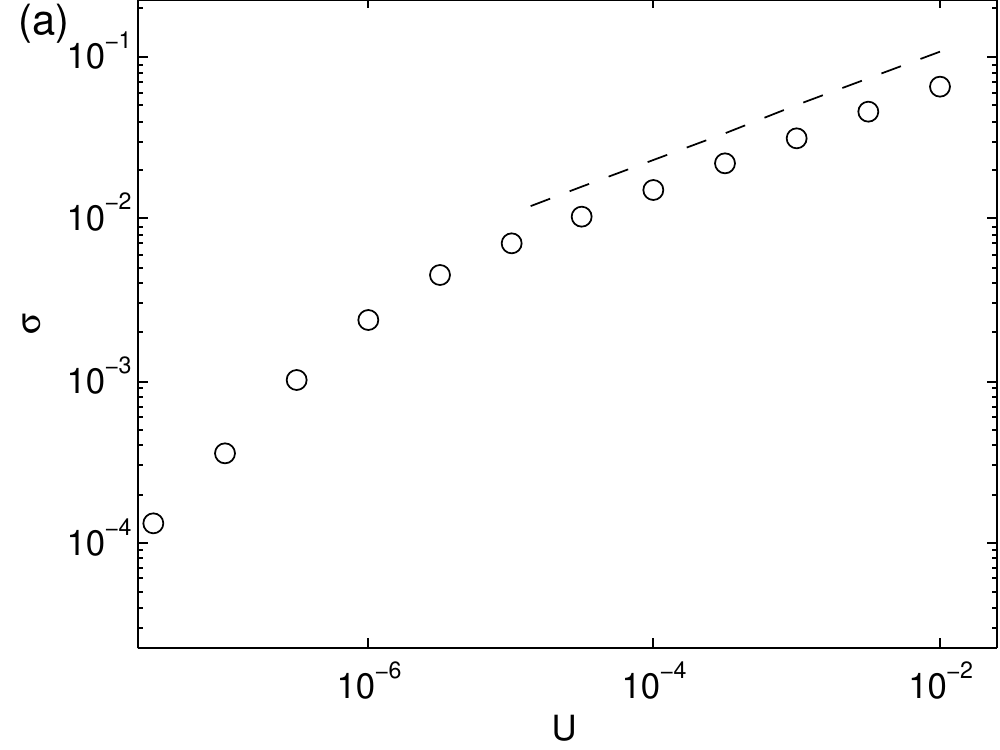}
\par\end{centering}

\begin{centering}
\includegraphics[width=1\columnwidth]{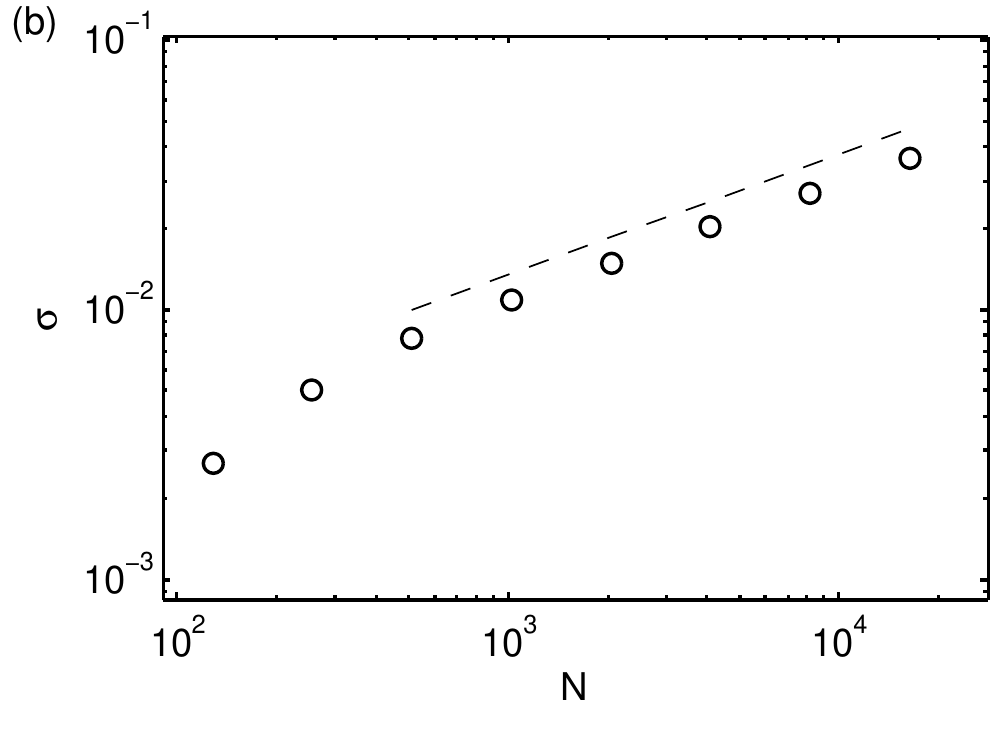}
\par\end{centering}

\caption{\label{fig:sd}Top: Stationary standard deviation $\lim_{t\to\infty}\sigma(t)$
versus mutation rate $U$ with $N=512$ and $s=0.01$ averaged over
$10^{5}$ mutations. The dashed line is $\sim U^{1/3}$ Bottom: width
increases with system size as a power law with a characteristic exponent,
with $U=10^{-5}$. The dashed line is $\sim N^{0.45}$.}
\end{figure}

From the simulation we also found the scaling of the steady-state
width. In the single-fixation regime we would expect $\sigma$ to
be small since most of the time the system has uniform fitness. Figure
\ref{fig:sd} shows $\sigma$ in multiple-mutation regime to be a
power law: $\sigma\sim U^{\eta}$ with $\eta=0.3-0.4$. The width
is also shown to go as $\sigma\sim N^{\alpha}$ and $\alpha$ is close
to $1/2$. 

Our model has a growth exponent close to $1/2$, which differs from
the linear Edwards-Wilkinson universality class ($\beta=1/4$) and
the non-linear KPZ universality class ($\beta=1/3$) \citep{Barabasi1995}.
However, we found that the growth of the width changes for larger
system sizes and longer time scales, which will be described in future
work.  

\section{\label{sec:Discussion}Discussion}

The model we study is restricted to one organism per site, as in \citep{Gordo2006},
in contrast to more detailed models divide a population into islands
or demes, each with a finite well-mixed population. However, the one
organism per site model may be approximated as a deme model with some
restrictions by grouping together sites into demes \citep{Korolev2010}.
The number of organisms per site affects the speed of the genetic
wave that sweeps each mutation to fixation. Since genetic stochasticity
is strong when there is one organism per site, the speed of the genetic
wave front is $v_{F}\sim s$, as in \citep{DOERING2003},\citep{Hallatschek2009}.
In the weak noise limit, when population in the demes are relatively
large and genetic drift is small, the genetic wave front moves with
the speed $v_{F}\sim\sqrt{s}$ \citep{Fisher1937}. Also, we have
not included a measure of the spatial dispersal of organisms after
reproduction. In any case, changing the genetic wave speed would change
$t_{\textrm{fix}}$, and therefore change the transition at which
interference occurs. Notably if $v_{F}\sim\sqrt{s}$, then $U_{\textrm{tr}}$
would depend on $s$. However, the fixation time dependence on the
total system size would not change since each island must be visited
sequentially regardless of wave speed. Therefore we expect our main
result that $R$ saturates with large $N$ will not change. Similarly,
generalizing $s$ to a distribution of beneficial effects, would affect
the time-scale of fixation, but we still expect $R$ to saturate with
large $N$. We confirmed this with an exponential distribution of
$s$ (not shown).

We have shown that spatial structure significantly affects the rate
of fixation in our model. In the multiple-mutation regime, the rate
of fixation is reduced and becomes independent of $N$ for large $N$,
similar to neutral fixations. Since beneficial mutations do not have
an overwhelming advantage over neutral and deleterious mutations,
it would be interesting to study a more general distribution of fitness
effects. It is possible for harmful mutations alone to accumulate
in one dimension when they are common enough \citep{Hallatschek2010}.

One dimensional populations have been used to model the frontiers
of expanding bacterial colonies, when the organisms at the frontier
reproduce faster than those in the interior \citep{Korolev2010},
although there are complications with boundary instabilities and boundary
growth. We hope that similar experiments could test our prediction
that the rate of fixation is independent of $N$.

A 2D version of our model could be representative of waves of spreading
mutations in an immobilized population. This would be interesting
to study since differences from the well-mixed model have already
been observed in simulation and experiment \citep{Campos2008,Gordo2006,Perfeito2006,Perfeito2008,Rosas2005},
and system size dependence has not been investigated. 
\begin{acknowledgments}
The authors thank Fereydoon Family, Ilya Nemenman, H.G.E Hentschel,
David Cutler, Joachim Krug, and Sorin Nicola-Tanase for helpful discussions.
We also thank the anonymous referees for many valuable comments. S.B.
acknowledges support from the US National Science Foundation through
Grant No. DMR-0812204.
\end{acknowledgments}
\bibliographystyle{apsrev}
\bibliography{article}

\begin{thebibliography}{36}
\expandafter\ifx\csname natexlab\endcsname\relax\def\natexlab#1{#1}\fi
\expandafter\ifx\csname bibnamefont\endcsname\relax
  \def\bibnamefont#1{#1}\fi
\expandafter\ifx\csname bibfnamefont\endcsname\relax
  \def\bibfnamefont#1{#1}\fi
\expandafter\ifx\csname citenamefont\endcsname\relax
  \def\citenamefont#1{#1}\fi
\expandafter\ifx\csname url\endcsname\relax
  \def\url#1{\texttt{#1}}\fi
\expandafter\ifx\csname urlprefix\endcsname\relax\def\urlprefix{URL }\fi
\providecommand{\bibinfo}[2]{#2}
\providecommand{\eprint}[2][]{\url{#2}}

\bibitem[{\citenamefont{Sniegowski and Gerrish}(2010)}]{Sniegowski2010}
\bibinfo{author}{\bibfnamefont{P.~D.} \bibnamefont{Sniegowski}}
  \bibnamefont{and} \bibinfo{author}{\bibfnamefont{P.~J.}
  \bibnamefont{Gerrish}}, \bibinfo{journal}{Philosophical transactions of the
  Royal Society of London. Series B, Biological sciences}
  \textbf{\bibinfo{volume}{365}}, \bibinfo{pages}{1255} (\bibinfo{year}{2010}).

\bibitem[{\citenamefont{Gerrish and Lenski}(1998)}]{Gerrish1998}
\bibinfo{author}{\bibfnamefont{P.}~\bibnamefont{Gerrish}} \bibnamefont{and}
  \bibinfo{author}{\bibfnamefont{R.}~\bibnamefont{Lenski}},
  \bibinfo{journal}{Genetica} \textbf{\bibinfo{volume}{102-103}},
  \bibinfo{pages}{127} (\bibinfo{year}{1998}).

\bibitem[{\citenamefont{Muller}(1932)}]{Muller1932}
\bibinfo{author}{\bibfnamefont{H.}~\bibnamefont{Muller}}, \bibinfo{journal}{Am.
  Nat.} \textbf{\bibinfo{volume}{66}}, \bibinfo{pages}{118}
  (\bibinfo{year}{1932}).

\bibitem[{\citenamefont{Fisher}(1930)}]{Fisher1930}
\bibinfo{author}{\bibfnamefont{R.}~\bibnamefont{Fisher}},
  \emph{\bibinfo{title}{{The Genetical Theory of Natural Selection}}}
  (\bibinfo{publisher}{Clarendon Press}, \bibinfo{address}{Oxford},
  \bibinfo{year}{1930}).

\bibitem[{\citenamefont{Hill and Robertson}(1966)}]{Hill1966}
\bibinfo{author}{\bibfnamefont{W.~G.} \bibnamefont{Hill}} \bibnamefont{and}
  \bibinfo{author}{\bibfnamefont{A.}~\bibnamefont{Robertson}},
  \bibinfo{journal}{Genetical Research} \textbf{\bibinfo{volume}{8}},
  \bibinfo{pages}{269} (\bibinfo{year}{1966}).

\bibitem[{\citenamefont{Barton}(1995)}]{Barton1995}
\bibinfo{author}{\bibfnamefont{N.~H.} \bibnamefont{Barton}},
  \bibinfo{journal}{Genetics} \textbf{\bibinfo{volume}{140}},
  \bibinfo{pages}{821} (\bibinfo{year}{1995}).

\bibitem[{\citenamefont{Kim and Orr}(2005)}]{Kim2005}
\bibinfo{author}{\bibfnamefont{Y.}~\bibnamefont{Kim}} \bibnamefont{and}
  \bibinfo{author}{\bibfnamefont{H.~A.} \bibnamefont{Orr}},
  \bibinfo{journal}{Genetics} \textbf{\bibinfo{volume}{171}},
  \bibinfo{pages}{1377} (\bibinfo{year}{2005}).

\bibitem[{\citenamefont{Barrett et~al.}(2006)\citenamefont{Barrett, M'Gonigle,
  and Otto}}]{Barrett2006}
\bibinfo{author}{\bibfnamefont{R.~D.~H.} \bibnamefont{Barrett}},
  \bibinfo{author}{\bibfnamefont{L.~K.} \bibnamefont{M'Gonigle}},
  \bibnamefont{and} \bibinfo{author}{\bibfnamefont{S.~P.} \bibnamefont{Otto}},
  \bibinfo{journal}{Genetics} \textbf{\bibinfo{volume}{174}},
  \bibinfo{pages}{2071} (\bibinfo{year}{2006}).

\bibitem[{\citenamefont{Desai and Fisher}(2007)}]{Desai2007}
\bibinfo{author}{\bibfnamefont{M.~M.} \bibnamefont{Desai}} \bibnamefont{and}
  \bibinfo{author}{\bibfnamefont{D.~S.} \bibnamefont{Fisher}},
  \bibinfo{journal}{Genetics} \textbf{\bibinfo{volume}{176}},
  \bibinfo{pages}{1759} (\bibinfo{year}{2007}).

\bibitem[{\citenamefont{Desai et~al.}(2007)\citenamefont{Desai, Fisher, and
  Murray}}]{Desai2007a}
\bibinfo{author}{\bibfnamefont{M.~M.} \bibnamefont{Desai}},
  \bibinfo{author}{\bibfnamefont{D.~S.} \bibnamefont{Fisher}},
  \bibnamefont{and} \bibinfo{author}{\bibfnamefont{A.~W.}
  \bibnamefont{Murray}}, \bibinfo{journal}{Current biology : CB}
  \textbf{\bibinfo{volume}{17}}, \bibinfo{pages}{385} (\bibinfo{year}{2007}).

\bibitem[{\citenamefont{Fogle et~al.}(2008)\citenamefont{Fogle, Nagle, and
  Desai}}]{Fogle2008}
\bibinfo{author}{\bibfnamefont{C.~A.} \bibnamefont{Fogle}},
  \bibinfo{author}{\bibfnamefont{J.~L.} \bibnamefont{Nagle}}, \bibnamefont{and}
  \bibinfo{author}{\bibfnamefont{M.~M.} \bibnamefont{Desai}},
  \bibinfo{journal}{Genetics} \textbf{\bibinfo{volume}{180}},
  \bibinfo{pages}{2163} (\bibinfo{year}{2008}).

\bibitem[{\citenamefont{Hallatschek}(2010)}]{Hallatschek2010a}
\bibinfo{author}{\bibfnamefont{O.}~\bibnamefont{Hallatschek}},
  \bibinfo{journal}{Proceedings of the National Academy of Sciences of the
  United States of America} \textbf{\bibinfo{volume}{108}},
  \bibinfo{pages}{1783} (\bibinfo{year}{2010}).

\bibitem[{\citenamefont{Park et~al.}(2010)\citenamefont{Park, Simon, and
  Krug}}]{Park2010}
\bibinfo{author}{\bibfnamefont{S.-C.} \bibnamefont{Park}},
  \bibinfo{author}{\bibfnamefont{D.}~\bibnamefont{Simon}}, \bibnamefont{and}
  \bibinfo{author}{\bibfnamefont{J.}~\bibnamefont{Krug}},
  \bibinfo{journal}{Journal of Statistical Physics}
  \textbf{\bibinfo{volume}{138}}, \bibinfo{pages}{381} (\bibinfo{year}{2010}).

\bibitem[{\citenamefont{Park and Krug}(2007)}]{Park2007}
\bibinfo{author}{\bibfnamefont{S.-C.} \bibnamefont{Park}} \bibnamefont{and}
  \bibinfo{author}{\bibfnamefont{J.}~\bibnamefont{Krug}},
  \bibinfo{journal}{Proceedings of the National Academy of Sciences of the
  United States of America} \textbf{\bibinfo{volume}{104}},
  \bibinfo{pages}{18135} (\bibinfo{year}{2007}).

\bibitem[{\citenamefont{Orr}(2010)}]{Orr2010}
\bibinfo{author}{\bibfnamefont{H.~A.} \bibnamefont{Orr}},
  \bibinfo{journal}{Philosophical transactions of the Royal Society of London.
  Series B, Biological sciences} \textbf{\bibinfo{volume}{365}},
  \bibinfo{pages}{1195} (\bibinfo{year}{2010}).

\bibitem[{\citenamefont{Martin and Lenormand}(2008)}]{Martin2008}
\bibinfo{author}{\bibfnamefont{G.}~\bibnamefont{Martin}} \bibnamefont{and}
  \bibinfo{author}{\bibfnamefont{T.}~\bibnamefont{Lenormand}},
  \bibinfo{journal}{Genetics} \textbf{\bibinfo{volume}{179}},
  \bibinfo{pages}{907} (\bibinfo{year}{2008}).

\bibitem[{\citenamefont{Maruyama}(1970)}]{Maruyama1970}
\bibinfo{author}{\bibfnamefont{T.}~\bibnamefont{Maruyama}},
  \bibinfo{journal}{Genetical Research} \textbf{\bibinfo{volume}{15}},
  \bibinfo{pages}{221} (\bibinfo{year}{1970}).

\bibitem[{\citenamefont{Lieberman et~al.}(2005)\citenamefont{Lieberman, Hauert,
  and Nowak}}]{Lieberman2005}
\bibinfo{author}{\bibfnamefont{E.}~\bibnamefont{Lieberman}},
  \bibinfo{author}{\bibfnamefont{C.}~\bibnamefont{Hauert}}, \bibnamefont{and}
  \bibinfo{author}{\bibfnamefont{M.}~\bibnamefont{Nowak}},
  \bibinfo{journal}{Nature} \textbf{\bibinfo{volume}{433}},
  \bibinfo{pages}{312} (\bibinfo{year}{2005}).

\bibitem[{\citenamefont{{Chaves Filho} et~al.}(2010)\citenamefont{{Chaves
  Filho}, de~Oliveira, and Campos}}]{ChavesFilho2010}
\bibinfo{author}{\bibfnamefont{V.~L.} \bibnamefont{{Chaves Filho}}},
  \bibinfo{author}{\bibfnamefont{V.~M.} \bibnamefont{de~Oliveira}},
  \bibnamefont{and} \bibinfo{author}{\bibfnamefont{P.~R.}
  \bibnamefont{Campos}}, \bibinfo{journal}{Physica A: Statistical Mechanics and
  its Applications} \textbf{\bibinfo{volume}{389}}, \bibinfo{pages}{5725}
  (\bibinfo{year}{2010}).

\bibitem[{\citenamefont{Whitlock}(2003)}]{Whitlock2003}
\bibinfo{author}{\bibfnamefont{M.~C.} \bibnamefont{Whitlock}},
  \bibinfo{journal}{Genetics} \textbf{\bibinfo{volume}{164}},
  \bibinfo{pages}{767} (\bibinfo{year}{2003}).

\bibitem[{\citenamefont{Habets et~al.}(2007)\citenamefont{Habets,
  Cz\'{a}r\'{a}n, Hoekstra, and de~Visser}}]{Habets2007}
\bibinfo{author}{\bibfnamefont{M.~G. J.~L.} \bibnamefont{Habets}},
  \bibinfo{author}{\bibfnamefont{T.}~\bibnamefont{Cz\'{a}r\'{a}n}},
  \bibinfo{author}{\bibfnamefont{R.~F.} \bibnamefont{Hoekstra}},
  \bibnamefont{and} \bibinfo{author}{\bibfnamefont{J.~A. G.~M.}
  \bibnamefont{de~Visser}}, \bibinfo{journal}{Proceedings. Biological sciences
  / The Royal Society} \textbf{\bibinfo{volume}{274}}, \bibinfo{pages}{2139}
  (\bibinfo{year}{2007}).

\bibitem[{\citenamefont{Gordo and Campos}(2006)}]{Gordo2006}
\bibinfo{author}{\bibfnamefont{I.}~\bibnamefont{Gordo}} \bibnamefont{and}
  \bibinfo{author}{\bibfnamefont{P.~R.~A.} \bibnamefont{Campos}},
  \bibinfo{journal}{Genetica} \textbf{\bibinfo{volume}{127}},
  \bibinfo{pages}{217} (\bibinfo{year}{2006}).

\bibitem[{\citenamefont{Perfeito et~al.}(2008)\citenamefont{Perfeito, Pereira,
  Campos, and Gordo}}]{Perfeito2008}
\bibinfo{author}{\bibfnamefont{L.}~\bibnamefont{Perfeito}},
  \bibinfo{author}{\bibfnamefont{M.~I.} \bibnamefont{Pereira}},
  \bibinfo{author}{\bibfnamefont{P.~R.~A.} \bibnamefont{Campos}},
  \bibnamefont{and} \bibinfo{author}{\bibfnamefont{I.}~\bibnamefont{Gordo}},
  \bibinfo{journal}{Biology letters} \textbf{\bibinfo{volume}{4}},
  \bibinfo{pages}{57} (\bibinfo{year}{2008}).

\bibitem[{\citenamefont{Hallatschek et~al.}(2007)\citenamefont{Hallatschek,
  Hersen, Ramanathan, and Nelson}}]{Hallatschek2007}
\bibinfo{author}{\bibfnamefont{O.}~\bibnamefont{Hallatschek}},
  \bibinfo{author}{\bibfnamefont{P.}~\bibnamefont{Hersen}},
  \bibinfo{author}{\bibfnamefont{S.}~\bibnamefont{Ramanathan}},
  \bibnamefont{and} \bibinfo{author}{\bibfnamefont{D.~R.}
  \bibnamefont{Nelson}}, \bibinfo{journal}{Proceedings of the National Academy
  of Sciences of the United States of America} \textbf{\bibinfo{volume}{104}},
  \bibinfo{pages}{19926} (\bibinfo{year}{2007}).

\bibitem[{\citenamefont{Hallatschek and Nelson}(2008)}]{Hallatschek2008}
\bibinfo{author}{\bibfnamefont{O.}~\bibnamefont{Hallatschek}} \bibnamefont{and}
  \bibinfo{author}{\bibfnamefont{D.~R.} \bibnamefont{Nelson}},
  \bibinfo{journal}{Theoretical population biology}
  \textbf{\bibinfo{volume}{73}}, \bibinfo{pages}{158} (\bibinfo{year}{2008}).

\bibitem[{\citenamefont{Hallatschek and Nelson}(2010)}]{Hallatschek2010}
\bibinfo{author}{\bibfnamefont{O.}~\bibnamefont{Hallatschek}} \bibnamefont{and}
  \bibinfo{author}{\bibfnamefont{D.~R.} \bibnamefont{Nelson}},
  \bibinfo{journal}{Evolution; international journal of organic evolution}
  \textbf{\bibinfo{volume}{64}}, \bibinfo{pages}{193} (\bibinfo{year}{2010}).

\bibitem[{\citenamefont{Korolev et~al.}(2010)\citenamefont{Korolev,
  Hallatschek, and Nelson}}]{Korolev2010}
\bibinfo{author}{\bibfnamefont{K.~S.} \bibnamefont{Korolev}},
  \bibinfo{author}{\bibfnamefont{O.}~\bibnamefont{Hallatschek}},
  \bibnamefont{and} \bibinfo{author}{\bibfnamefont{D.~R.}
  \bibnamefont{Nelson}}, \bibinfo{journal}{Reviews of Modern Physics}
  \textbf{\bibinfo{volume}{82}}, \bibinfo{pages}{1691} (\bibinfo{year}{2010}).

\bibitem[{\citenamefont{Kimura}(1962)}]{KIMURA1962}
\bibinfo{author}{\bibfnamefont{M.}~\bibnamefont{Kimura}},
  \bibinfo{journal}{Genetics} \textbf{\bibinfo{volume}{47}},
  \bibinfo{pages}{713} (\bibinfo{year}{1962}).

\bibitem[{\citenamefont{Barab\'{a}si and Stanley}(1995)}]{Barabasi1995}
\bibinfo{author}{\bibfnamefont{A.-L.} \bibnamefont{Barab\'{a}si}}
  \bibnamefont{and} \bibinfo{author}{\bibfnamefont{H.~E.}
  \bibnamefont{Stanley}}, \emph{\bibinfo{title}{{Fractal concepts in surface
  growth}}} (\bibinfo{publisher}{Cambridge University Press},
  \bibinfo{year}{1995}), ISBN \bibinfo{isbn}{0521483182}.

\bibitem[{\citenamefont{Family and Vicsek}(1985)}]{Family1985}
\bibinfo{author}{\bibfnamefont{F.}~\bibnamefont{Family}} \bibnamefont{and}
  \bibinfo{author}{\bibfnamefont{T.}~\bibnamefont{Vicsek}},
  \bibinfo{journal}{Journal of Physics A: Mathematical and General}
  \textbf{\bibinfo{volume}{18}}, \bibinfo{pages}{L75} (\bibinfo{year}{1985}).

\bibitem[{\citenamefont{Doering et~al.}(2003)\citenamefont{Doering, Mueller,
  and Smereka}}]{DOERING2003}
\bibinfo{author}{\bibfnamefont{C.}~\bibnamefont{Doering}},
  \bibinfo{author}{\bibfnamefont{C.}~\bibnamefont{Mueller}}, \bibnamefont{and}
  \bibinfo{author}{\bibfnamefont{P.}~\bibnamefont{Smereka}},
  \bibinfo{journal}{Physica A: Statistical Mechanics and its Applications}
  \textbf{\bibinfo{volume}{325}}, \bibinfo{pages}{243} (\bibinfo{year}{2003}).

\bibitem[{\citenamefont{Hallatschek and Korolev}(2009)}]{Hallatschek2009}
\bibinfo{author}{\bibfnamefont{O.}~\bibnamefont{Hallatschek}} \bibnamefont{and}
  \bibinfo{author}{\bibfnamefont{K.}~\bibnamefont{Korolev}},
  \bibinfo{journal}{Physical Review Letters} \textbf{\bibinfo{volume}{103}},
  \bibinfo{pages}{108103} (\bibinfo{year}{2009}).

\bibitem[{\citenamefont{Fisher}(1937)}]{Fisher1937}
\bibinfo{author}{\bibfnamefont{R.~A.} \bibnamefont{Fisher}},
  \bibinfo{journal}{Annals of Eugenics} \textbf{\bibinfo{volume}{7}},
  \bibinfo{pages}{355} (\bibinfo{year}{1937}).

\bibitem[{\citenamefont{Campos et~al.}(2008)\citenamefont{Campos, Neto,
  de~Oliveira, and Gordo}}]{Campos2008}
\bibinfo{author}{\bibfnamefont{P.~R.~A.} \bibnamefont{Campos}},
  \bibinfo{author}{\bibfnamefont{P.~S. C.~A.} \bibnamefont{Neto}},
  \bibinfo{author}{\bibfnamefont{V.~M.} \bibnamefont{de~Oliveira}},
  \bibnamefont{and} \bibinfo{author}{\bibfnamefont{I.}~\bibnamefont{Gordo}},
  \bibinfo{journal}{Evolution; international journal of organic evolution}
  \textbf{\bibinfo{volume}{62}}, \bibinfo{pages}{1390} (\bibinfo{year}{2008}).

\bibitem[{\citenamefont{Perfeito et~al.}(2006)\citenamefont{Perfeito, Gordo,
  and Campos}}]{Perfeito2006}
\bibinfo{author}{\bibfnamefont{L.}~\bibnamefont{Perfeito}},
  \bibinfo{author}{\bibfnamefont{I.}~\bibnamefont{Gordo}}, \bibnamefont{and}
  \bibinfo{author}{\bibfnamefont{P.~R.} \bibnamefont{Campos}},
  \bibinfo{journal}{The European Physical Journal B}
  \textbf{\bibinfo{volume}{51}}, \bibinfo{pages}{301} (\bibinfo{year}{2006}).

\bibitem[{\citenamefont{Rosas et~al.}(2005)\citenamefont{Rosas, Gordo, and
  Campos}}]{Rosas2005}
\bibinfo{author}{\bibfnamefont{A.}~\bibnamefont{Rosas}},
  \bibinfo{author}{\bibfnamefont{I.}~\bibnamefont{Gordo}}, \bibnamefont{and}
  \bibinfo{author}{\bibfnamefont{P.}~\bibnamefont{Campos}},
  \bibinfo{journal}{Physical Review E} \textbf{\bibinfo{volume}{72}},
  \bibinfo{pages}{012901} (\bibinfo{year}{2005}).

\end{thebibliography}

\end{document}